\documentclass[12pt,a4paper]{article}

\usepackage{dcolumn}
\usepackage{bm}
\usepackage{epsfig}

\newcommand{\half}{\frac{1}{2}}

\newcommand{\beq}{\begin{equation}}
\newcommand{\eeq}{\end{equation}}
\newcommand{\bea}{\begin{eqnarray}}
\newcommand{\eea}{\end{eqnarray}}

\newcommand{\dd}{\partial}

\newcommand{\av}[1]{\left\langle #1 \right\rangle}

\newcommand{\RL}{R_L}

\newcommand{\eqnr}[1]{Eq.(\ref{#1})}

\newcommand{\vep}{\varepsilon}

\newcommand{\cdim}{C_{\vep}}
\newcommand{\cdiminf}{C_{\vep,\infty}}

\title{An exact expression for the Reynolds number dependence of the
energy dissipation rate in homogeneous, isotropic turbulence}
\author{W. David McComb, Arjun Berera, Matthew Salewski and Samuel Yoffe \\
School of Physics and Astronomy, \\
University of Edinburgh, UK.\\
Email: wdm@ph.ed.ac.uk}

\begin{document}
\maketitle

\begin{abstract}
The Reynolds number dependence of the dimensionless dissipation rate
$\cdim=\vep L/U^3$ is derived directly from the Karman-Howarth equation as
$
\cdim = \cdiminf[1 +  A/R_L]
$,
where the coefficient $A$ depends on the second- and
third-order structure functions, $R_L = UL/\nu$, $L$ is the
integral length scale, and $U$ is the rms velocity. Fitting this
form to results from DNS indicates that $A$ is effectively constant and
hence the predicted dependence of $\cdim$ on Reynolds number is as $R_L^{-1}$.
\end{abstract}


\maketitle

In recent years, there has been great interest in the Reynolds number
dependence of the dissipation rate in homogeneous, isotropic fluid turbulence: see
\cite{Sreenivasan84}\nocite{Sreenivasan98}\nocite{Kaneda03}\nocite{Burattini05}\nocite{Pearson02}\nocite{Pearson04a}\nocite{Donzis05}\nocite{Seoud07}\nocite{Bos07}\nocite{Mazellier08}-\cite{Goto09}.
Apart from its intrinsic fundamental interest, this is a key factor in
the free decay of turbulence, which is one of the most studied aspects
of the turbulence problem: see, for example,
\cite{George92}\nocite{Speziale92}\nocite{Skrbek00}-\cite{Wang02}, and
the many references therein.

We consider the dimensionless dissipation rate 
\beq
\varepsilon = \cdim U^3/L,
\label{taylordiss}
\eeq
which was put forward in 1935 by Taylor \cite{Taylor35} on the basis of
dimensional arguments. Here $U$ is the rms velocity of the fluid and $L$
is the integral length scale.
As early as 1953, Batchelor \cite{Batchelor71} (in the first edition of
this book) presented evidence to suggest that $\cdim$ tends to a constant with
increasing Reynolds number. However, the present interest in the subject
stems from the seminal paper by Sreenivasan \cite{Sreenivasan84}, who
established that in grid turbulence $\cdim$ became constant for
Taylor-Reynolds numbers greater than about 50 but observed that the
actual value could depend on the flow configuration and initial
conditions.  

Attempts have been made to establish a theoretical relationship between
the dissipation rate and the Reynolds number. Lohse \cite{Lohse94} used
a mean-field closure of the Karman-Howarth equation to obtain an
approximate form, whereas Doering and Foias \cite{Doering02} have
established upper and lower bounds to be satisfied by such
relationships. In this Letter, we derive a new and exact relationship
between the dissipation and the Reynolds number. We begin by stating the
Karman-Howarth equation and reviewing the relevant phenomenology. 

As is well known, the Karman-Howarth equation \cite{VonKarman38} is
derived directly from the Navier-Stokes equation and is an exact
relationship expressing conservation of energy. It may be written in
terms of structure functions as \cite{Landau59}:
\beq
-\frac{2}{3}\varepsilon - \half \frac{\dd S_2}{\dd t} = \frac{1}{6
r^4}\frac{\dd}{\dd r}(r^4  S_3) - \frac{\nu}{r^4}\frac{\dd}{\dd
r}\left(r^4 \frac{\dd S_2}{\dd r}\right),
\label{khe}
\eeq
where the structure function of order $n$ is given by
\beq
S_n = \av{\left(u(r) -u(0)\right)^n}.
\eeq
We begin by re-writing it as an expression for the dissipation rate,
thus: 
\beq
\varepsilon =   - \frac{3}{4} \frac{\dd S_2}{\dd t}  - \frac{1}{4
r^4}\frac{\dd}{\dd r}(r^4  S_3) + \frac{3\nu}{2r^4}\frac{\dd}{\dd
r}\left(r^4 \frac{\dd S_2}{\dd r}\right).
\label{khgendiss}
\eeq

Now we re-visit the Richardson-Kolmogorov phenomenology of this equation
\cite{Batchelor71}, as this will be of specific use to us later on. We
first consider the stationary case and, as usual, this requires the
injection of energy at scales $r\geq r_I$ (say). As the Karman-Howarth
equation is local in $r$, we restrict our attention to scales less than
the injection scale $r_I$ and set the time-derivative equal to zero.
With these steps, equation (\ref{khgendiss}) becomes:
\beq
\varepsilon = - \frac{1}{4 r^4}\frac{\dd}{\dd r}(r^4  S_3) + \frac{3\nu}{2r^4}\frac{\dd}{\dd
r}\left(r^4 \frac{\dd S_2}{\dd r}\right).
\label{khdiss}
\eeq
We note that each term on the right hand side is separately a function
of $r$ but that jointly they are constant, as required to match the
dissipation rate on the left hand side. If we go further, and restrict
the values of the scale to $r\geq r_d$, where $r_d$ is the dissipation
lengthscale, then for sufficiently large values of the Reynolds number
the well known arguments of Kolmogorov (K41)
\cite{Kolmogorov41a,Kolmogorov41b} apply. In order to apply this
rigorously, we take the infinite Reynolds number limit. This means
taking the limit $\nu \rightarrow 0$ such that $\vep$ remains constant
\cite{Batchelor71}. The viscous term may then be set equal to zero, and the
remaining term on the right hand side represents the inertial flux of
energy, which is now constant, and equal to the dissipation rate. Hence:
\beq
\varepsilon = - \frac{1}{4 r^4}\frac{\dd}{\dd r}(r^4  S_3).
\label{khkoldiss}
\eeq
From this equation it is a simple matter to recover Kolmogorov's
famous `4/5' law for the third-order structure function, thus:
\beq
S_3(r)= -\frac{4}{5}\varepsilon r.
\label{fourfive}
\eeq

Next we make a \emph{change of variables} based on constant velocity- and
length-scales, $V$ and $l$, respectively. As an example, we will put the
second-order structure function into a dimensionless form. We do this by
expanding $S_2$ in a Taylor-Maclaurin series. Note that this is a general
operation and is not restricted to any particular values of $r$. Then,
introducing a dimensionless variable $x$, through $r=lx$, we have
\bea
S_2(r) & = & S_2(0)  + S'_2(0)r + \frac{1}{2}S_2''(0)r^2 + \dots, \nonumber \\
       & = & S_2(0)  + S'_2(0)lx + \frac{1}{2}S_2''(0)l^2x^2 + \dots, \nonumber \\
       & = & V^2\left[f_2(0)  + f'_2(0)x + \frac{1}{2}f_2''(0)x^2 + \dots\right], \nonumber\\
       & \equiv & V^2 f_2(x),
\eea
where, in the third line, we have renamed coefficients in order to
extract a factor $V^2$, and also to absorb powers of $l$.  Note that the
primes denote differentiation with respect to $r$ in the first two lines
but with respect to $x$ in the third.  Lastly, we have defined the
dimensionless second-order structure function $f_2(x)$ as the sum of the
Maclaurin series in the new coefficients. 

We can repeat this process for structure functions of any order $n$, and we
summarise the general result as:
\beq
S_n = V^n f_n(x), \qquad \mbox{with} \qquad x=\frac{r}{l}. 
\label{dimless}
\eeq
We should note that this procedure involves no approximations or non-trivial
assumptions. It merely introduces the $f_n$ as dimensionless forms of the 
structure functions. As they are dimensionless, their dependence on $r$
must be scaled by some length, here denoted by $l$. 

Then we make the specific choices $V = U$, the root-mean-square velocity,
and $l=L$, the integral length scale. With these choices, and substituting from
(\ref{dimless}), we change equation (\ref{khdiss}) to the form:
\beq
\varepsilon = \frac{A_3 U^3}{L} + \frac{A_2\nu U^2}{L^2},
\label{interdiss}
\eeq
where the coefficients $A_3$ and $A_2$ are given by
\beq
A_3 \equiv -\frac{1}{4x^4}\frac{\dd}{\dd x}\left[x^4 f_3(x)\right],
\label{a3coefft}
\eeq
and
\beq
A_2 \equiv \frac{3}{2x^4}\frac{\dd}{\dd x}\left[x^4 \frac{\dd f_2}{\dd x}\right].
\label{a2coefft}
\eeq
Then, if we divide both sides of (\ref{interdiss}) by $U^3/L$ we may
write this as
\beq
\cdim = A_3 +  \frac{A_2}{R_L} = A_3\left[1 + \frac{A_2}{A_3 R_L}\right],
\label{cdim}
\eeq
where the dimensionless dissipation $\cdim$ is as defined in
(\ref{taylordiss}) and the Reynolds number is given by $R_L = U L /
\nu$. Note that the $f_n$ are determined by (\ref{dimless}), along with
this choice of scaling variables, and hence so also are the coefficients
$A_3$ and $A_2$. We also note that this equation is still just the
Karman-Howarth equation: no approximation has been made. 

Let us consider the asymptotic behaviour of this expression for
large Reynolds numbers. In general the coefficients $A_2$ and $A_3$ may
depend on the Reynolds number (although, as we shall see later,
comparison with the results from DNS suggests that they are essentially
constant, albeit with possibly some dependence on initial conditions).
However, the asymptotic properties of (\ref{cdim}) must be those of the
Karman-Howarth equation. So taking the limit of infinite Reynolds
numbers, and comparing (\ref{cdim}) with (\ref{khdiss}), we note that
the coefficient $A_3$ becomes constant, while the term $A_2/R_L$
vanishes. Then we obtain, by analogy with (\ref{khkoldiss}), the
asymptotic form of (\ref{cdim}) as
\beq
\lim_{\RL\rightarrow \infty} \cdim \equiv   \cdiminf = A_3.
\label{cdiminf}
\eeq
Substituting back into (\ref{cdim}) then yields
\beq
\cdim = \cdiminf \left[1 +  A/R_L \right],
\label{cdimfinal}
\eeq
where
\beq
A= A_2/A_3.
\eeq

We now discuss the extension of our results to freely decaying isotropic
turbulence. Of course, the neglect of the time-derivative term in
(\ref{khe}) is quite usual, even for decaying turbulence, provided that
the Reynolds number is large enough and that one restricts attention to
the inertial range. For instance, this step is required in order to
derive the `4/5' law for decaying turbulence and is known as \emph{local
stationarity}. However, we wish to consider a more general approach in
which we introduce the time dependence of all statistical quantities, so
that we now have $\varepsilon(t)$ and $S_n(r,t)$, at any time $t$.

In order to make comparisons with the stationary case, we take some
fiducial time $t=t_e$, when the turbulence is assumed to have evolved
from arbitrary initial conditions to a state determined solely by the
Navier-Stokes equations. Then, as before, we introduce a change of
variables, such that
\beq
S_n(r,t) = U^n(t_e) g_n(x,\tau),
\label{tdimless}
\eeq
where now
\beq
x=r/L(t_e);   \qquad  \tau = t/T; \qquad T=L(t_e)/U(t_e).
\eeq
Just as before, in the stationary case, the introduction of the
dimensionless (but now time-dependent) structure functions $g_n(x,\tau)$
may be accomplished by the use of Taylor series; although, in this case,
it is for a function of two variables. 

It should be emphasised that the form (\ref{tdimless}) is \emph{not} a
similarity solution of any kind. If we were to drop the dependence on
the variable $\tau$ then this would amount to an assumption of
self-preservation, as introduced by von Karman in 1938
\cite{VonKarman38}. However, this is one of the vexed issues of
turbulence theory. It runs into difficulties, not least because during
free decay the characteristic length changes from the integral length
scale to the viscous scale in the final period of the decay: see, for
instance, \cite{Speziale92}. As it is, we reiterate that we do not make
any such assumption and that we retain the full time dependence of the
problem.

With this in mind, it is easily shown that substituting (\ref{tdimless})
into (\ref{khgendiss}) leads to a generalization of equation
(\ref{cdim}) to the form:
\beq
\cdim = (A_3 - B_2)\left[1 +
\frac{1}{R_L}\frac{A_2}{A_3 - B_2}\right],
\label{finaldisstime}
\eeq
where the coefficients $A_2$ and $A_3$ are still defined by equations
(\ref{a2coefft}) and (\ref{a3coefft}), but now with $f$ replaced by $g$,
and the new coefficient $B_2$ is given by
\beq
B_2 \equiv \left. \frac{3}{4}\frac{\dd g_2}{\dd \tau}\right|_{\tau_e}.
\eeq

Let us now consider the effect of including this time dependence. We
introduce the inertial flux which we denote by $\Pi_{max}$. (Formally,
this quantity is defined in wavenumber space and for our present
purposes we are interested in its maximum value with respect to
wavenumber: for a discussion see \cite{McComb90a}.) Then, as is well
known, for increasing Reynolds number, the maximum flux approaches the
dissipation rate from below; or:
\beq
\lim_{\RL \rightarrow \infty} \Pi_{max} \rightarrow \vep
\label{statinv},
\eeq
for stationary turbulence. Our present analysis indicates that this
cannot, \emph{in principle}, be the case for freely decaying turbulence.

We may see this as follows. Rewriting the coefficient $B_2$ in terms of
the structure function, we have
\beq 
B_2 = \left.\frac{3}{4}\frac{\dd g_2}{\dd \tau}\right|_{\tau_e} =
\left.\frac{3}{4}\frac{L(t_e)}{U^3(t_e)}\frac{\dd S_2}{\dd t}\right|_{t_e}. 
\eeq
For free decay, this coefficient must be negative. Taking its modulus we have
\beq
\lim_{\RL \rightarrow \infty} \vep\rightarrow \frac{U^3}{L}(A_3 + |B_2|).
\eeq 
Thus, in freely decaying turbulence, we have
\beq
\lim_{\RL \rightarrow \infty} \frac{\Pi_{max}}{\vep}\rightarrow \frac{A_3}{A_3 +
|B_2|} < 1,
\eeq
and so the inertial flux of energy never \emph{quite} reaches the same
value as the dissipation. This result has implications for the
interpretation of the Kolmogorov (K41) picture, when a comparison is
made of forced and decaying homogeneous, isotropic turbulence, and we
shall investigate this further in future work.

For completeness, we now give the extension of equation (\ref{cdim}) to
the case of free decay, as:
\beq
\cdim = \cdiminf \left[1 +  A / R_L\right],
\label{cdimdec}
\eeq
where the dimensionless dissipation at infinite Reynolds number and the
coefficient $A$ are now given by
\beq
\cdiminf = A_3 - B_2 \quad \mbox{and} \quad A= A_2/(A_3 - B_2).
\label{cdimdecinf}
\eeq

We may compare our results to those of other theories. Lohse
\cite{Lohse94} used a mean-field approximation to the Karman-Howarth
equation and obtained the asymptotic result $\cdiminf = (6/b)^{3/2}$,
where $b$ is the prefactor in the Kolmogorov inertial-range form of the
second-order structure function. This may be compared with our equation
(\ref{cdiminf}), which in contrast gives $\cdiminf = A_3$. This
difference from our result is only to be expected because a closure
invariably involves expressing $S_3$ in terms of $S_2$. (It should be
emphasised that our remarks here have no implications for the validity
or accuracy of Lohse's approximation.)

It is also of interest to compare equation (\ref{interdiss}), which is
an intermediate stage in our calculation, to the result for an upper
bound on the dissipation as given by Doering and Foias \cite{Doering02}.
This is featured in their abstract as
\[
\varepsilon \leq c_1 \nu \frac{U^2}{l^2} +c_2 \frac{U^3}{l},
\]
and corresponds to their equation (40). Here the coefficients $c_1$ and
$c_2$ depend on the shape of the forcing function, while $l$ is its
longest length-scale. Obviously this is quite different from our own
result, where the corresponding parameters depend on the fluid
turbulence and not on the forcing. Also, we have an \emph{equality}, rather than
an \emph{inequality}. Nevertheless, it can easily be shown that the two results
are equivalent and this will be presented later, in a fuller account of
this work.

\begin{figure}[!h]
\epsfig{figure=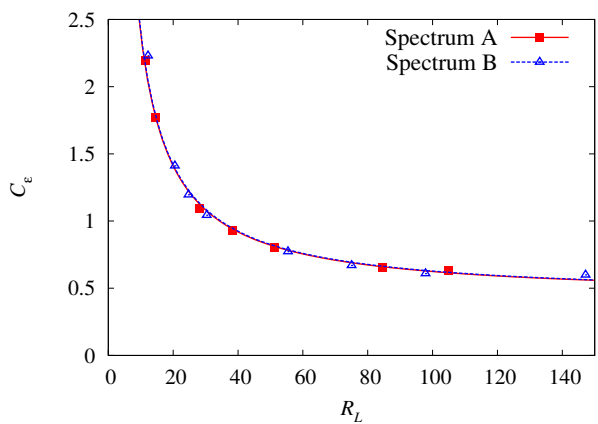,height=6cm}
\caption{Values of $\cdim$ from DNS for two different initial spectra
plotted against the integral length scale Reynolds number, $\mathcal{R}_L$. These
points were fitted to the curve $y(x)=a+(b/x)$, as given by Eq.
(\ref{cdimdec}).}
\label{fig}
\end{figure}

Lastly, we illustrate the dependence of the dissipation on Reynolds
number by fitting the general form of equation (\ref{cdimdec}) to
the results of direct numerical simulations (DNS) of the Navier-Stokes
equation. This is shown for two different initial spectra, which are
generated using the following equation:
\begin{equation}\label{inispec}
E(k,0) = C_1\mathrm{k}^{C_2}\exp \Big\{-C_3\mathrm{k}^{C_4}\Big\},
\end{equation}
where the values taken for the parameters $C_1$-$C_4$ are as given in
Table~\ref{ispectra_table}. 
\begin{table}
\begin{tabular}{l|c|c|c|c}
\hline
Initial spectra & $C_1$ & $C_2$ & $C_3$ & $C_4$ \\
\hline
Spectrum A & 0.0017 & 4 &      0.08 & 2 \\
Spectrum B & 0.08 & 2 &  0.0824 & 2 \\
\hline
\end{tabular}
\caption{Initial energy spectra parameter values for use in the
numerical computations. These parameters are substituted into
\eqnr{inispec} to generate the required initial
spectrum.}\label{ispectra_table}
\end{table}
Note that Spectrum A and Spectrum B have low-wavenumber regions varying
as $k^4$ and $k^2$, respectively. The fitting process gives asymptotic
values of $\cdiminf = 0.43\pm 0.01$ and $\cdiminf = 0.43\pm 0.02$,
respectively.

Figure \ref{fig} illustrates the way in which our new exact relationship
between the dimensionless dissipation rate and the Reynolds number can
be used to fit the data generated by numerical or other experiments. In
particular it supports the idea that the coefficient $A$ is constant
with respect to the Reynolds number, even at low Reynolds numbers. 

Summing up, we have taken the Karman-Howarth equation as an expression
for the dissipation rate and transformed it into a more useful form, by
making a change of variables. Then we have used its known properties to
identifiy the asymptotic dimensionless dissipation coefficient and put
it into the yet more useful form of (\ref{cdimfinal}). At no time have
we made any approximation or similarity assumption. Our only
mathematical assumption is the underlying one of theoretical physics:
that all variables corresponding to physical observables are
mathematically well-behaved. From our comparison with DNS, we conclude
that the low-Reynolds number dependence of the dissipation coefficient
is of the form: $\mbox{constant} \times \RL^{-1}$. 

Currently we are working to extend the simulations to forced turbulence,
in order to study the stationary case. In particular, we intend to study
the detailed behaviour of the coefficients at small Reynolds numbers,
including the extent of their dependence on the initial conditions. This
work will be the subject of a full report in due course.

\section*{}

AB and SY were funded by the STFC, while WDM wishes to acknowledge the
support provided by the award of an Emeritus Fellowship by the
Leverhulme Trust, along with the  hospitality of the Isaac Newton
Institute (in the form of a Visiting Fellowship) and that of the
Institute for Mathematical Sciences, Imperial College. He would also
like to thank Francis Barnes, Claude Cambon, Stuart Coleman, Gregory
Falkovich, Jorgen Frederiksen, Katepalli R. Sreenivasan, and Christos
Vassilicos for stimulating discussions, helpful comments or other
assistance.



\end{document}